\documentstyle[seceq,epsf,wrapft,preprint]{ptptex}
\newcommand{\Hess}{{\rm Hess}\,}
\newcommand{\Tr}{{\rm Tr}\,}

\renewcommand{\Re}{{\rm Re}\,}
\renewcommand{\Im}{{\rm Im}\,}

\title{%        %You can use \\ for explicit line-break
Ghost orbit bifurcations in semiclassical spectra
}
\author{%       %Use \sc for the family name
Thomas {\sc Bartsch},
J\"org {\sc Main} and G\"unter {\sc Wunner}
}

\inst{%         %Affiliation, neglected when [addenda] or [errata]
Institut f\"ur Theoretische Physik 1,
Universit\"at Stuttgart,\\
D-70550 Stuttgart, Germany}
\recdate{%      %Editorial Office will fill in this.
\today
}

\abst{%       %this abstract is neglected when [addenda] or [errata]
Gutzwiller's semiclassical trace formula for the density of states in a 
chaotic system diverges near bifurcations of periodic orbits, where it must
be replaced with uniform approximations. It is well known that, when 
applying these approximations, complex predecessors of orbits created in the
bifurcation (``ghost orbits'') can produce clear signatures in the
semiclassical spectra. We demonstrate that these orbits themselves can 
undergo bifurcations, resulting in complex, non-generic bifurcation 
scenarios. We do so by studying an example taken from the Diamagnetic Kepler 
Problem. By application of normal form theory, we construct an analytic
description of the complete bifurcation scenario, which is then used to 
calculate the pertinent uniform approximation. The ghost orbit bifurcation 
turns out to produce signatures in the semiclassical spectrum in much the same
way as a bifurcation of real orbits would.
}

\begin{document}

\maketitle

\section{Introduction}

Since its discovery in the early 1970s, Gutzwiller's trace formula
\cite{Gut67} has become a widely used tool for the interpretation of
quantum spectra of systems whose classical counterpart exhibits chaotic
behaviour. It represents the density of states of the quantum system as a
sum over a smooth part and fluctuations from all periodic orbits of the
classical system, which are all calculated from purely classical data. This
formula assumes all periodic orbits to be isolated, so that it fails close
to a bifurcation of periodic orbits, where different orbits approach one
another arbitrarily closely, leading to a divergence of the periodic orbit
contributions.

This failure can be overcome with the help of \emph{uniform approximations}
\cite{Alm87,Sie96,Sch97a,Sie98} which take into account the contributions
of all bifurcating orbits collectively. Although in generic Hamiltonian
systems only codimension-one bifurcations can be observed as a single
control parameter is varied, in practical applications of uniform
approximations bifurcation scenarios of higher codimensions must be
included:\cite{Mai97,Sch97b} \ If a periodic orbit successively undergoes
several bifurcations, uniform approximations capable of collectively
treating all orbits participating in the whole complicated bifurcation
scenario are needed.

In the construction of most uniform approximations known, \emph{ghost
orbits} \cite{Kus93} in the complexified classical phase space play a
crucial role, because real periodic orbits that are born in a bifurcation
tend to have ghost orbit predecessors before they turn real. So far, these
ghost orbits have never been observed to undergo bifurcations themselves,
although there is no a priori reason why this should be impossible. In
fact, we will now present an example of a ghost orbit bifurcation which
occurs in connection with a generic period-quadrupling bifurcation of a
real orbit. We will show that the construction of a uniform approximation
requires the inclusion of this bifurcation and that traditional normal form
theory can be extended to also cover bifurcations of ghost orbits. A more
detailed presentation of our results can be found in \citen{Bar1,Bar2}.

\section{The bifurcation scenario}
\label{BifSec}
As an example, we study the hydrogen atom in a magnetic field, which
has been described in detail, e.g., in Refs.\
\citen{Fri89,Has89,Wat93}.  We assume the nucleus fixed and regard the
electron as a structureless point charge. Due to a scaling property of the
Hamiltonian, the classical dynamics does not depend on the energy $E$ and
the magnetic field strength $\gamma$ separately, but only on the scaled
energy $\tilde E = \gamma^{-2/3} E$. To plot periodic orbits, we use
scaled semiparabolical coordinates $\mu^2 = \tilde r + \tilde z,\ \nu^2 =
\tilde r - \tilde z$.

To look for ghost orbits, we complexify the classical phase space by
allowing coordinates and momenta to assume complex values. Since our
Hamiltonian is real, the system is symmetric with respect to complex
conjugation. Therefore, the complex conjugate of a periodic orbit is a
periodic orbit itself, so that orbits usually occur in complex conjugate
pairs. In exceptional cases, however, a periodic ghost orbit can coincide
with its complex conjugate. These symmetric ghost orbits then have real
periods and actions.

\begin{wrapfigure}{r}{\halftext}
  \epsfxsize = \halftext
  \epsfbox{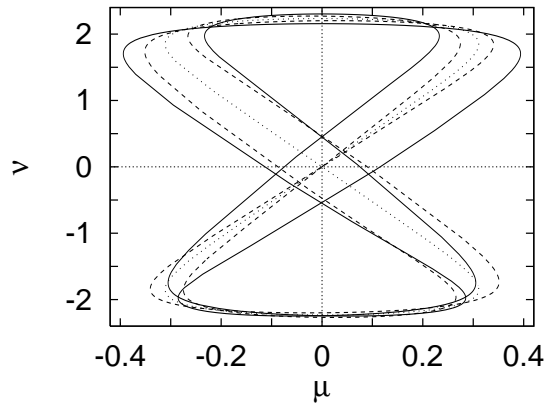}
  \caption{Real orbits at scaled energy $\tilde E = -0.340 > \tilde E_c$,
  drawn in scaled semiparabolical coordinates. Solid and dashed curves:
  stable and unstable satellite orbits. Dotted curve: balloon orbit}
  \label{realOrb}
\end{wrapfigure}

One of the shortest periodic orbits of the diamagnetic Kepler problem is
the balloon orbit, which we will now focus attention on. It undergoes a
period-quadrupling at a scaled energy of (in scaled atomic units, which we
use throughout) $\tilde E_c = -0.342\,025$. If $\tilde E>\tilde E_c$, a
stable and an unstable real satellite orbit of quadruple period exist. At
$\tilde E_c$, they simultaneously collide with the balloon orbit and
vanish. The real orbits are shown in figure \ref{realOrb} at the scaled
energy of $\tilde E = -0.340$. The solid and dashed curves represent the
stable and unstable satellite orbits, respectively. For comparison, the
balloon orbit is shown as a dotted curve. If $\tilde E<\tilde E_c$, no real
satellites are present, but there are a stable and an unstable ghost orbit
instead, both of which are symmetric with respect to complex conjugation.
Thus, at $\tilde E_c$ a generic island-chain bifurcation takes place.

\begin{figure}
  \parbox{\halftext}{
    \epsfxsize = \halftext
    \epsfbox{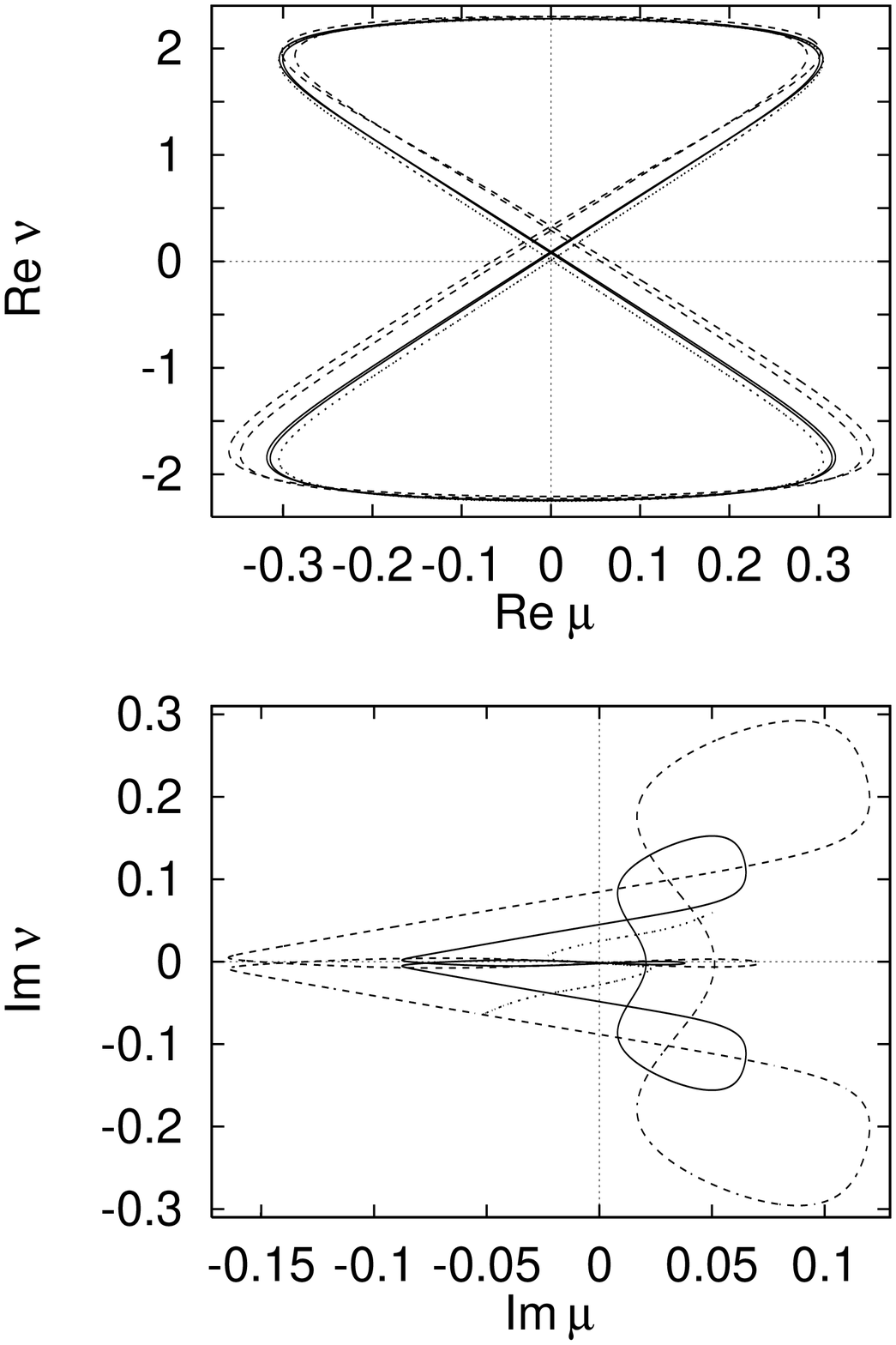}
    \caption{Real and imaginary parts of complex ghost orbits at scaled
      energy $\tilde E = -0.343$. Solid and dotted curves: stable and
      unstable ghost satellites created in the period-quadrupling
      bifurcation of the balloon orbit at $\tilde E_c =
      -0.342\,025$. Dashed curve: additional ghost orbit created in the
      ghost bifurcation at $\tilde E_c' = -0.343\,605$.}
    \label{Kplx343}}
  \hfill
  \parbox{\halftext}{
    \epsfxsize = \halftext
    \epsfbox{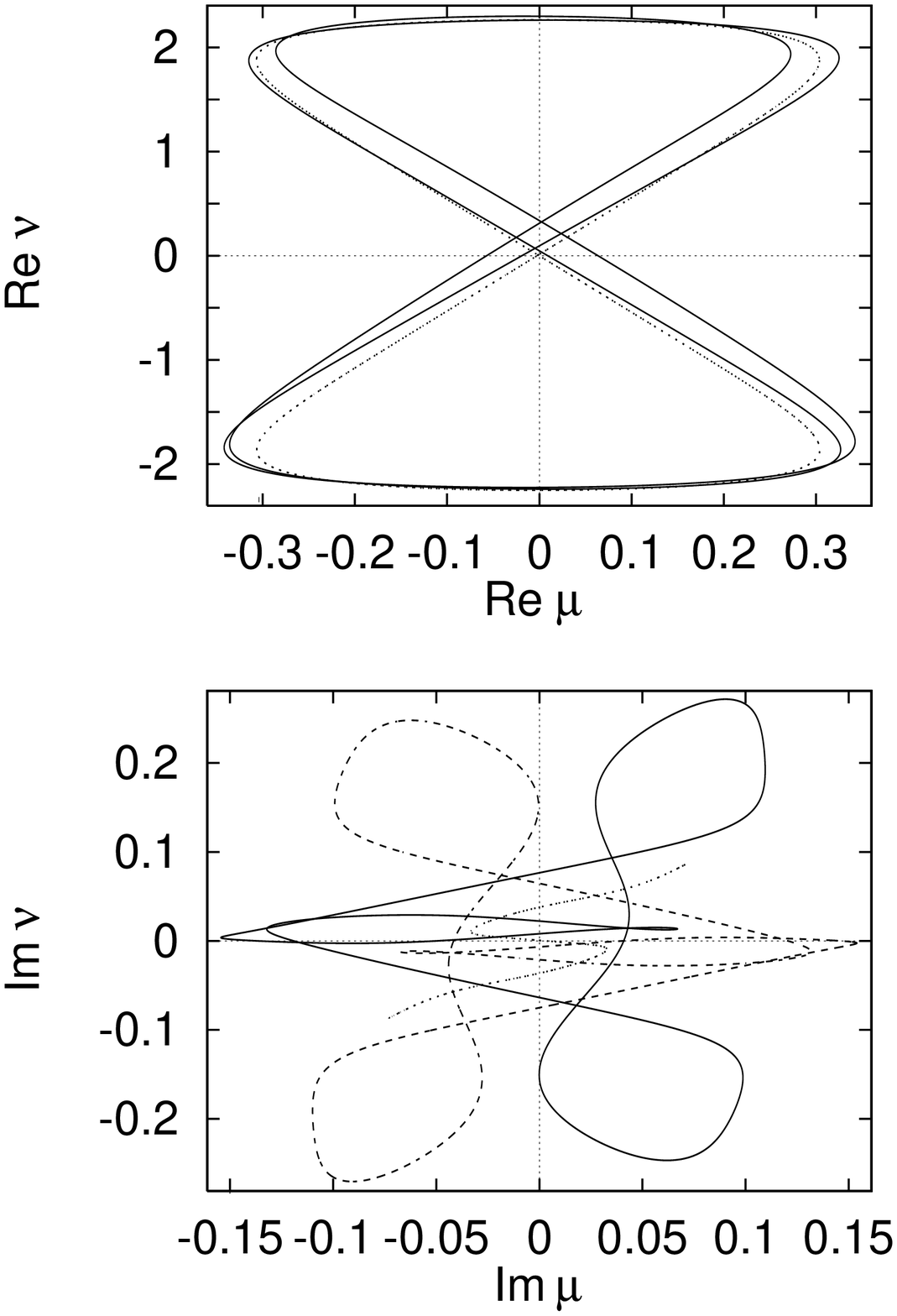}
    \caption{Ghost orbits at scaled energy $\tilde E= -0.344 < \tilde
      E_c'$. Solid and dashed curves: asymmeric ghost orbits (real parts
      coincide) created at the ghost bifurcation at $\tilde E_c' =
      -0.343\,605$. Dotted curve: unstable ghost satellite created at the
      period-quadrupling bifurcation of the balloon orbit}
    \label{Kplx344}}
\end{figure}

Furthermore, there is an additional ghost orbit in the phase space close to
this bifurcation that is also symmetric with respect to complex
conjugation. At $\tilde E_c' = -0.343\,605 < \tilde E_c$, this orbit
collides with the stable ghost satellite, the two orbits loose their
symmetry and become a pair of complex conjugate ghosts at $\tilde E<\tilde
E_c'$. For scaled energies above and below $\tilde E_c'$, the ghost orbits
are depicted in figures \ref{Kplx343} and \ref{Kplx344}, respectively. From
the imaginary parts it can clearly be seen that the symmetry with respect
to complex conjugation is lost in the bifurcation.

This is the first example of a ghost orbit bifurcation that has been
described in the literature. Its occurrence presents an additional
challenge to the construction of a uniform approximation because 
an approximation that deals with the generic period quadrupling bifurcation
only  diverges at the bifurcation energy of the ghost orbit bifurcation
and becomes undefined below $\tilde E_c'$ where the stable ghost satellite
orbit used in its construction does not exist any more.

\begin{figure}
  \parbox{\halftext}{
    \epsfxsize = \halftext
    \epsfbox{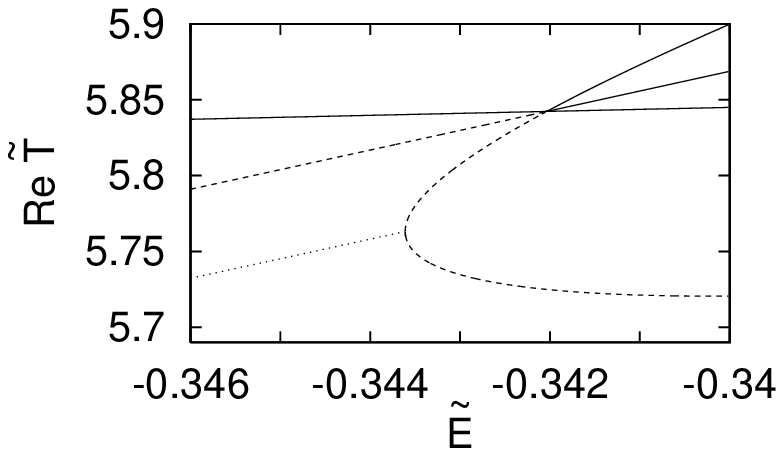}}
  \hfill
  \parbox{\halftext}{
    \epsfxsize = \halftext
    \epsfbox{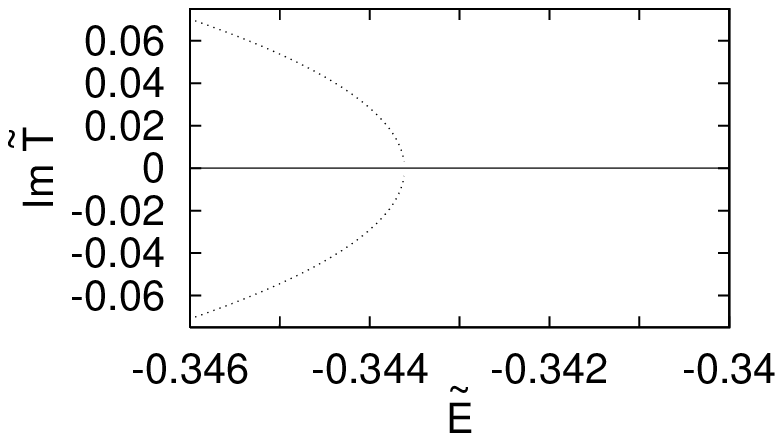}}
  \caption{Orbital periods of the orbits involved in the bifurcations as
    functions of the scaled energy $\tilde E = E \gamma^{-2/3}$}
  \label{orbDat}
\end{figure}

All periodic orbit parameters required for the construction of the uniform
approximation were calculated numerically. The complete numerical data is
described in detail in \citen{Bar1,Bar2}. As an example, the orbital
periods are shown in figure \ref{orbDat}. From the successive confluences
of the periods, the sequence of bifurcations becomes clearly visible: The
period of four repetitions of the balloon orbit, which is always real, is
indicated by a nearly horizontal line at $\tilde T \approx 5.84$. Above
$\tilde E_c$, there are two additional solid curves representing the
periods of the stable (upper curve) and unstable (lower curve) real
satellite orbits. At $\tilde E_c$, these curves change from solid to dashed
as the satellite orbits become ghosts. Below $\tilde E_c$, the unstable
ghost satellite does not undergo any further bifurcations in the energy
range shown, whereas the stable satellite collides, at $\tilde E_c'$, with
the additional ghost orbit. The latter can clearly be seen not to be
involved in the bifurcation at $\tilde E_c$. Below $\tilde E_c'$, these two
orbits are complex conjugates of each other. Thus, the real parts of their
periods coincide, whereas the imaginary parts are different from zero and
have opposite signs.

\section{The general form of the uniform approximation}
\label{UnifSec}

Before we return to classical normal form theory in section \ref{NFSec},  
we introduce, in this section,
the basic formulas for the quantum density
of states necessary for the construction of the uniform semiclassical
approximation.
The density of states of a quantum system with the Hamiltonian $\hat H$
can be expressed with the help of the Green's function $G(E)=(E-\hat
H)^{-1}$ as
\begin{equation}
  d(E) = -\frac{1}{\pi}\,\Im \Tr \, G(E) \; ,
\end{equation}
where the trace of the Green's function can be evaluated in the
coordinate re\-pre\-sen\-ta\-tion,
\begin{equation}
   \Tr \, G(E)
 = \int d\mib{x'}d\mib{x}\,\delta(\mib{x'}-\mib{x})\,
    G(\mib{x'}\mib{x},E) \; .
\end{equation}
The first step in the formulation of {\em periodic orbit theory}
\cite{Gut67} is to replace the Green's function
$G(\mib{x'}\mib{x},E)$ with its semiclassical Van Vleck-Gutzwiller
approximation.  For systems with two degrees of freedom the
semiclassical approximation to the Green's function reads
\begin{equation}
   G(\mib{x'}\mib{x}, E) = \frac{1}{i\hbar\sqrt{2\pi i\hbar}}
     \sum_{\rm class.~traj.}\sqrt{D}
     \exp\left\{\frac{i}{\hbar}S(\mib{x'}\mib{x}, E)
                 -i\frac{\pi}{2}\nu\right\} \; .
\end{equation}
Here, the sum extends over all classical trajectories running from
$\mib{x}$ to $\mib{x'}$ at energy $E$, $S$ is the action of a trajectory,
$\nu$ its Maslov index, and $D$ is given in terms of second derivatives of
the action. The contribution of a single orbit to the trace can be
evaluated by introducing coordinates parallel and perpendicular to the
orbit.  The integration along the orbit can then be performed in a
straightforward fashion. If one notes that on a periodic orbit the action
$S(\mib{x'}\mib{x}, E)$ is stationary with respect to the transverse
coordinates, Gutzwiller's trace formula for isolated periodic orbits is
finally obtained by integrating over the transverse coordinates using the
stationary-phase approximation.  It is this last step which fails close to
a bifurcation, where periodic orbits are not isolated.

The basic idea of the stationary-phase approximation is to approximate
the action function by a quadratic function in the neighbourhood of any
individual periodic orbit. To achieve a collective treatment of the
bifurcating orbits, we need to find an ansatz function $\Phi$ on a
Poincar\'e surface of section which has got stationary points corresponding
to all periodic orbits to be included in the uniform approximation. We can
then relate $\Phi$ to the classical action function $S(\mib{x'}\mib{x}, E)$
by means of a suitable, albeit unknown, coordinate transformation. In terms
of $\Phi$, the uniform approximation can be shown to assume the form
\cite{Sie96}
\begin{equation}\label{unifApprox}
 d(E) = \frac{1}{2\pi^2m\hbar^2}\, \Re 
         \exp\left\{\frac{i}{\hbar}S_0(E)
                    -i\frac{\pi}{2}\hat{\nu}\right\}
         \int dY\,dP_Y'\, X(Y, P_Y')\,
	    \exp\left\{\frac{i}{\hbar}\Phi(Y,P_Y')\right\}\;.
\end{equation}
The coefficient $X$ can
be evaluated at the stationary points of $\Phi$, where it assumes the value
\begin{equation}
  X \stackrel{\rm sp}{=} 
         \frac{\{m\}T}{\sqrt{|\Tr M-2|}}\sqrt{|\Hess\Phi|}\;.
\end{equation}
Here, $T$ and $M$ denote the period and the monodromy matrix of the
corresponding classical orbit, $\Hess\Phi$ the Hessian determinant of the
normal form, and the notation $\{m\}$ is meant to
indicate that this factor does not occur at the satellite orbits.

\section{Normal-form description of the bifurcation} 
\label{NFSec}
A systematic means to construct an ansatz function $\Phi$ is provided by
normal form theory.\cite{Bir27,Gus66} \ Here, we adopt a normal form used
by Schomerus \cite{Sch98} to describe codimension-two bifurcations. In
canonical polar coordinates $(I,\varphi)$, which are connected to Cartesian
coordinates $(p,q)$ by
\begin{equation}\label{polKoord}
  p = \sqrt{2I}\,\cos\varphi\;,\qquad q=\sqrt{2I}\,\sin\varphi\;,
\end{equation}
it reads
\begin{equation}\label{NF}
  \Phi = \varepsilon I + a I^2 + bI^2\cos(4\varphi)
        + cI^3(1+\cos(4\varphi))\;.
\end{equation}
This normal form turns out to qualitatively describe the sequence of
bifurcations encountered here for suitably chosen parameter values.

\begin{wrapfigure}{r}{\halftext}
  \epsfxsize = \halftext
  \epsfbox{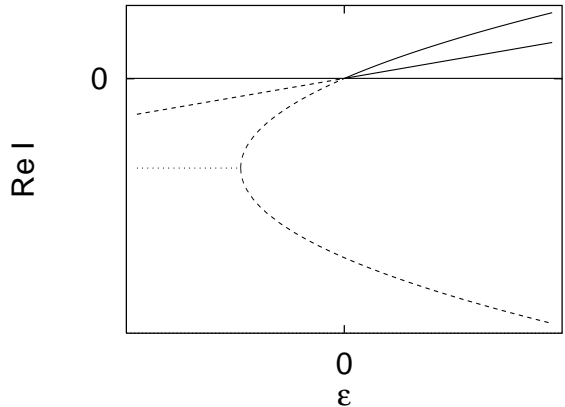}
  \caption{A sketch of the bifurcation scenario given by the normal form
    (\ref{NF}) for the case $|a|>|b|$ and $c<0$. Solid curves: real
    orbits. Dashed curves: ghost orbits symmetric with respect to complex
    conjugation. Dotted curve: a pair of complex conjugate ghosts.}
  \label{IEtaFig}
\end{wrapfigure}

To establish the connection to the classical bifurcation scenario, we have
to determine the stationary points of $\Phi$. The central periodic orbit at
$I=0$ does not show up as a stationary point because the polar coordiante
system (\ref{polKoord}) is singular there. In the case $|a|>|b|$ and $c<0$
we find three stationary points, each of which appears at four different
angles $\varphi$ because it corresponds to an orbit of quadruple
period. For these stationary points, the real parts of the action
coordinates $I$ are shown in figure \ref{IEtaFig}. The dotted line in this
figure corresponds to a pair of complex conjugate stationary points.

To interprete these results, we observe that according to its
definition (\ref{polKoord}) the coordinate $I$ is positive for real
orbits and that the action $\Phi(I,\varphi)$ is real for
real $I, \varphi$. Therefore, negative real solutions $I$ correspond
to ghost orbits which are symmetric with respect to complex
conjugation and thus have real actions, whereas a complex $I$ indicates
an asymmetric ghost orbit.

If $\varepsilon > 0$, we have two stationary points at positive values of
$I$ and one stationary point at a negative $I$. They correspond to two real
satellite orbits of quadruple period and a symmetric ghost orbit. As
$\varepsilon$ decreases through zero, two stationary points simultaneously
move from positive to negative $I$, thus indicating that the two real
satellites become ghosts in an island-chain bifurcation. Finally, two
stationary points with negative action coordinates collide at a negative
value of $\varepsilon$ and become complex. This describes a ghost orbit
bifurcation in which two symmetric ghost orbits collide and loose their
symmetry.

The normal form (\ref{NF}) thus qualitatively reproduces the bifurcation
scenario described above. The normal form parameters $\varepsilon, a,b,c$
have to be determined so as to make the description quantitatively
correct. To this end we calculate the stationary values of the normal form,
equate them to the actions of the periodic orbits and then solve for the
normal form parameters. As can be shown by a lengthy
calculation,\cite{Bar2} \ these parameters are uniquely determined by the
actions and the energy if we choose $\varepsilon = \tilde E - \tilde E_c$
and require all parameters to depend continuously on the energy.

\section{Evaluation of the uniform approximation}
\label{EvalSec}
After the ansatz function $\Phi$ has been completely specified, a
suitable approximation to the coefficient $X$ in (\ref{unifApprox})
remains to be found. We assume $X$ to be independent of $\varphi$,
and as the value of $X$ is known at the stationary points of $\Phi$ at
four different values of $I$ (including $I=0$), we approximate $X$ by
the third order polynomial $p(I)$ interpolating between the four given
points. This choice ensures that our approximation reproduces
Gutzwiller's isolated-orbits formula if, sufficiently far away from
the bifurcations, we evaluate the integral in
stationary-phase-approximation.
Thus, the uniform approximation takes its final form
\begin{equation}\label{UnifSol}
 d(E) = \frac{1}{2\pi^2m\hbar^2}\,\Re
     \exp\left\{\frac{i}{\hbar}S_0(E)
                          -i\frac{\pi}{2}\hat{\nu}\right\}
     \int dY\,dP_Y'\,p(I)\,
           \exp\left\{\frac{i}{\hbar}\Phi(Y,P_Y')\right\}\;,
\end{equation}
which contains known functions only and can be evaluated numerically.

\begin{figure}
  \parbox{\halftext}{
    \epsfxsize = \halftext
    \epsfbox{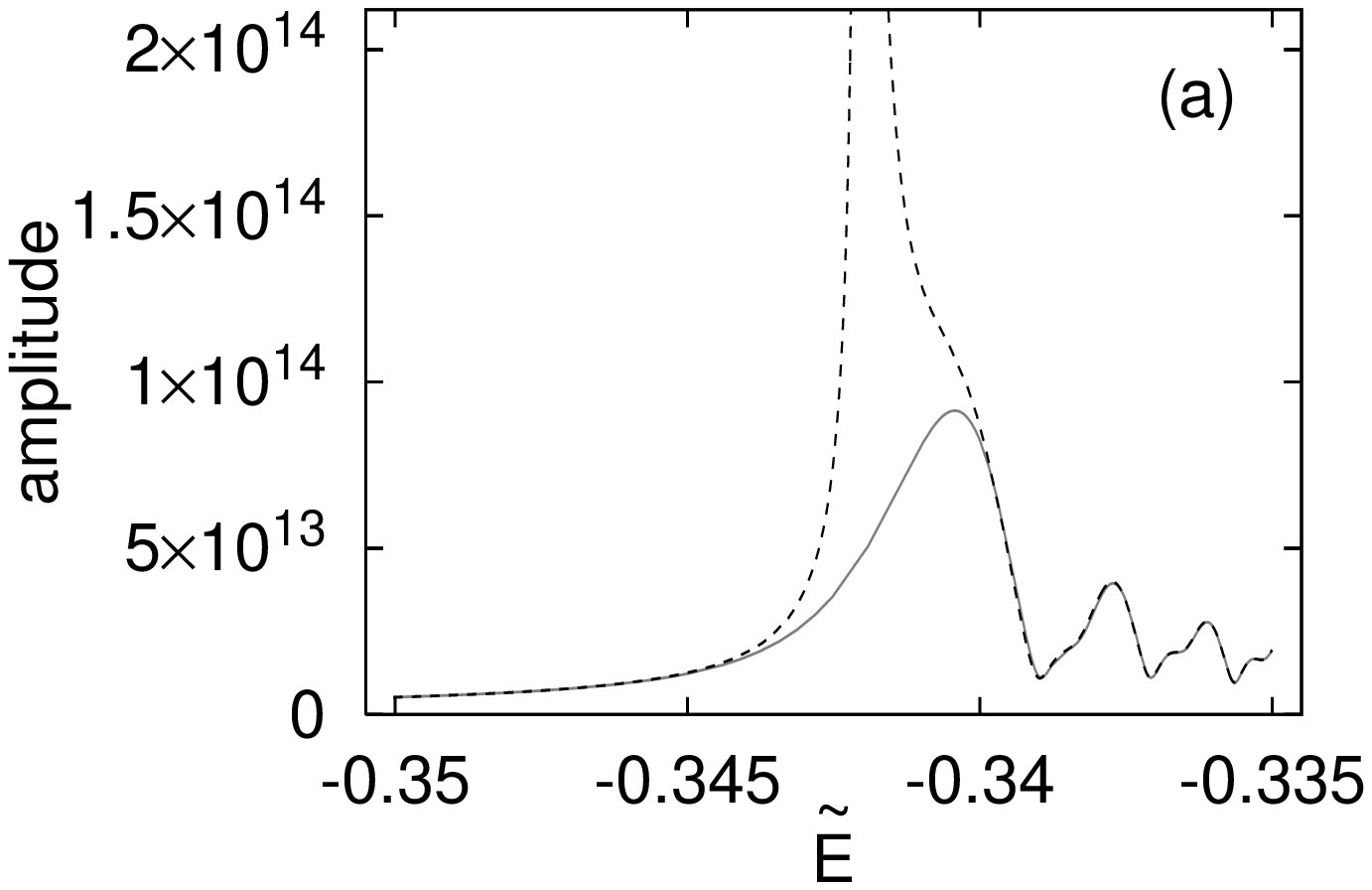}}
  \hfill
  \parbox{\halftext}{
    \epsfxsize = \halftext
    \epsfbox{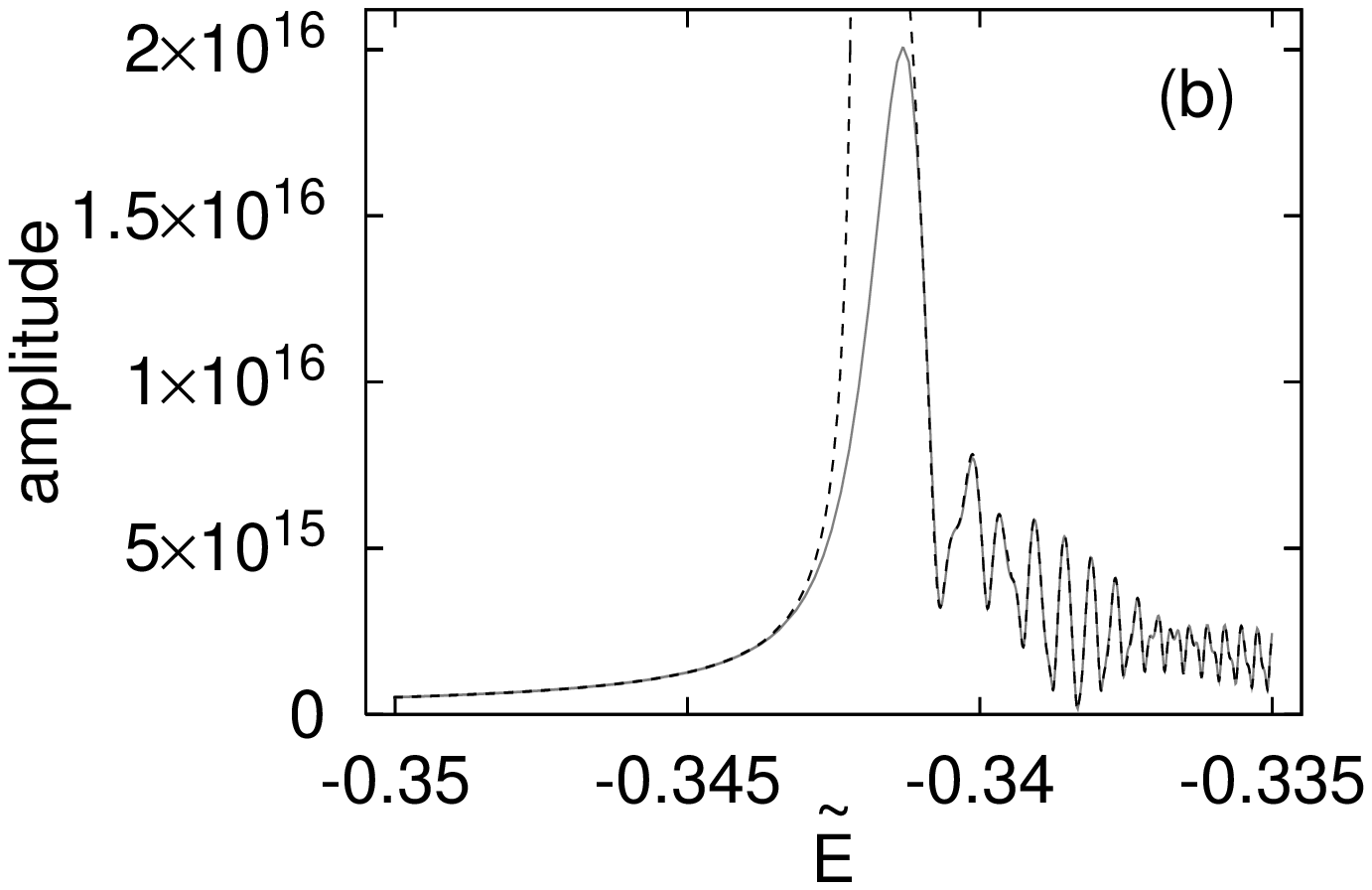}}
  \caption{Uniform approximation to the contribution of the considered
    bifurcations to the density of states for two different values of the
    magnetic field strength: $(a) \gamma = 10^{-12}$, $(b) \gamma =
    10^{-14}$. Solid curves: uniform approximations. Dashed curves:
    Gutzwiller's trace formula.}
  \label{UnifFig}
\end{figure}

We calculated the uniform approximation (\ref{UnifSol}) for two different
values of the magnetic field strength $\gamma$. The results are shown in
figure \ref{UnifFig}. To suppress the highly oscillatory contributions
originating from the factor $\exp\left\{\frac{i}{\hbar}S_0(E)\right\}$, we
plot the absolute value of (\ref{UnifSol}) instead of the real part.  As
can be seen, the uniform approximation proposed is finite at the
bifurcation energies, and, as the distance from the bifurcations increases,
asymptotically goes over into the results of Gutzwiller's trace
formula. Even the complicated oscillatory structures in the density of
states caused by interferences between the contributions from the different
real orbits involved at $\tilde{E}>\tilde{E}_c=-0.342\,025$ are perfectly
reproduced by our uniform approximation. We also see that the higher the
magnetic field strength, the farther away from the bifurcation the
asymptotic (Gutzwiller) behaviour is acquired. The magnetic field
dependence of the transition into the asymptotic regime can be traced back
to the fact that, due to the scaling properties of our system, the scaling
parameter $\gamma^{1/3}$ plays the r\^ole of an effective Planck's
constant, therefore the lower $\gamma$ becomes, the more accurate the
semiclassical approximation will be.

\section{Conclusion}

We have shown that in Hamiltonian systems with mixed regular-chaotic
dynamics {\em ghost orbit bifurcations} can occur besides the bifurcations
of real orbits. These are of special importance when they appear in the
vicinity of bifurcations of real orbits, since they turn out to produce
signatures in the semiclassical spectra much the same as those of the real
orbits. Consequently, the traditional theory of uniform approximations for
bifurcations of real orbits must be extended to also include the effects of
bifurcating ghost orbits.

We have illustrated the phenomenon of bifurcating ghost orbits in the
neighbourhood of bifurcations of real orbits by way of example for the
period-quadrupling of the balloon orbit in the diamagnetic Kepler problem,
and have demonstrated how normal form theory can be extended for this case
so as to allow for a unified description of both real {\em and} complex
bifurcations.

We picked the example mainly for its simplicity, since (a) the real orbit
considered is one of the shortest fundamental periodic orbits in the
diamagnetic Kepler problem and (b) the period-quadrupling is the lowest
period-$m$-tupling possible ($m=4$) that exhibits the island-chain
bifurcation typical of all higher $m$. Thus we expect ghost orbit
bifurcations to appear also for longer-period orbits, and, in particular,
in the vicinity of all higher period-$m$-tupling bifurcations of real
orbits.

In fact, a general discussion\cite{Bar2} of the bifurcation scenarios
described by the normal form (\ref{NF}) and a more general variant of its
for different values of the parameters leads us to the conclusion that the
appearance of ghost orbit bifurcations in the vicinity of bifurcating real
orbits is the rule, rather than the exception, in general systems with
mixed regular-chaotic systems, and thus one of their generic features.
% It will be
%interesting and rewarding to study higher period-$m$-tuplings with respect
%to the appearance of ghost orbit bifurcations, and to extend ordinary
%normal form theory to also include the contributions of ghost orbit
%bifurcations for all higher $m$.

\end{document}